\documentclass[english,a4paper]{article}
\usepackage[T1]{fontenc}
\usepackage[latin9]{inputenc}
\usepackage{booktabs}
\usepackage{url}
\usepackage{amsmath}
\usepackage{amssymb}
\usepackage{graphicx}
\usepackage{esint}
\usepackage{rotating}
\usepackage{authblk}

\makeatletter

\providecommand{\tabularnewline}{\\}

\makeatother

\usepackage{babel}

\title{A Mathematical Model of Flavescence Dorée Epidemiology}

\author[1]{Federico Lessio}
\author[1]{Alessandro Portaluri}
\author[2,3]{Francesco Paparella\thanks{francesco.paparella@unisalento.it}}
\author[1]{Alberto Alma}

\affil[1]{Department of Agricultural, Forest and Food
Sciences,  University of Torino, Italy.}
\affil[2]{Dipartimento di Matematica e Fisica
``Ennio De Giorgi'', University of Salento, Italy.}
\affil[3]{INFN sez. di Lecce, Italy.}

\begin{document}

\maketitle
\begin{abstract}
Flavescence dorée (FD) is a disease of gra\-pe\-vine transmitted
by an insect vector, \emph{Scaphoideus titanus} Ball. At present,
no prophylaxis exists, so mandatory control procedures (e.g. removal
of infected plants, and insecticidal sprays to avoid transmission)
are in place in Italy and other European countries. We propose a model
of the epidemiology of FD by taking into account the different aspects
involved into the transmission process (acquisition of the disease,
latency and expression of symptoms, recovery rate, removal and replacement
of infected plants, insecticidal treatments, and the effect of hotbeds).
The model was constructed as a system of first order nonlinear ODEs
in four compartment variables. We perform a bifurcation analysis of
the equilibria of the model using the severity of the hotbeds as the
control parameter. Depending on the non-dimensional grapevine density
of the vineyard we find either a single family of equilibria in which
the health of the vineyard gradually deteriorates for progressively
more severe hotbeds, or multiple equilibria that give rise to sudden
transitions from a nearly healthy vineyard to a severely deteriorated
one when the severity of the hotbeds crosses a critical value. These
results suggest some lines of intervention for limiting the spread
of the disease.\\
{\bf keywords}{Flavescence dorée \and Grapevine epidemiology \and Critical transition
\and Fold bifurcation \and \emph{Scaphoideus titanus} Ball}
\end{abstract}

\section{Introduction}

Flavescence dorée (hereafter FD) is a serious disease of grapevine,
wi\-de\-spread in many European countries, caused by phytoplasmas
belonging to 16SrV-C and
 16SrV-D ribosomal groups \cite{Malembic-Maher 2009}. Symptoms of
FD include leaf yellowing or redness, lack of lignification of canes,
lack of blossom, and so on. The infected plants stop to produce grapes,
and die after a few years. Symptoms of FD are usually shown after
a latency period of 1-3 years from infection; young plants are more
likely to show symptoms just one year after infection \cite{Morone 2007,Osler 2002}. 

FD is transmitted vine-to-vine by an insect vector, \emph{Scaphoideus
titanus} Ball (Hemiptera: Cicadellidae), native to North America and
introduced into Europe in the late 1950s \cite{Bonfils and Schvester 1960,Chuche14}.
\emph{S. titanus} feeds and reproduces only on grapevine (\emph{Vitis}
spp.), has a single generation per year, and overwinters in the egg
stage, laid under the bark of grapes \cite{Vidano 1964,Chuche14}.
Eggs start to hatch during spring, and the insect over-goes through
five nymphal instars before becoming adult during summer \cite{Vidano 1964,Bressan 2006}.
Nymphs from the 3rd and later instars acquire phytoplasmas when feeding
on infected plants, and after a latency period lasting 4-5 weeks (meanwhile
becoming adults) they are able to inoculate phytoplasmas to healthy
plants \cite{Bressan 2005}. Once infective, insects retain vector
capability through their lifetime; on the other hand, no transovarial
transmission has been proved for \emph{S. titanus} at present, therefore
newly born insects have to feed on infected plants in order to acquire
phytoplasmas \cite{Alma 1997}. Other insects are acknowledged to
be occasional vectors \cite{Filippin 2009}, however their role in
the spread of Flavescence dorée to date is not considered to be important. 

Infected plants may be subject to recovery, with symptoms disappearing
within a few years after the infection with recovery rates that depend
on cultivar and age of plants, the youngest being the less able to
recover \cite{Morone 2007}. Observed recovery rates are highly variable
and range from 1\% to 70\% of the infected plants, but show a strong
inverse dependence on the abundance of the vector insect, with the
highest recovery rates observed in vineyards subject to aggressive
insecticide treatments, and the lowest in vineyards subject to no
treatments \cite{Morone 2007,Zorloni 2008}. This supports the notion
that recovered plants are not immune from reinfection. However, recovered
plants are not a source of phytoplasmas for insects \cite{Galetto 2009}. 

In Italy FD is subject to mandatory control procedures, including
sprays of insecticide against the vector and removal of the infected
plants, which, however, may have higher cost than insecticide treatment.
In many vine-growing areas abandoned vineyards and woods containing
wild grapevine act as hotbeds of both phytoplasmas and \emph{S. titanus}
\cite{Pavan 2012,Lessio 2014}. Adults of \emph{S. titanus} are able
to move from untreated to treated vineyards up to a distance of about
300 m \cite{Lessio 2014}.

In this paper we model the dynamics of the spread of FD over time,
by considering the different aspects involved into (or influencing
the) transmission process. The model could be used for forecasting
the epidemiology of FD in a vineyard, given the knowledge of some
parameters. More importantly, it highlights the key ecological factors
involved in the infection process, and thus it offers guidance for
planning an adequate response.

From a mathematical point of view, the model is a system of nonlinear,
first-order, ordinary differential equations in the compartments $S$,
$L$, $I$, $G$, modelling the dynamical behaviour of healthy full-grown,
latent, infected and young (nursery) plants, respectively. 

The rest of the paper is structured as follows. In Section \ref{sec:The-model}
we formulate the model and present the main mathematical results.
In Section \ref{sec:Discussion} we discuss the ecological significance
of those results. Conclusions are given in Section \ref{sec:Conclusions}.
The Appendix \ref{sec:Appendix} contains proofs and other mathematical
details that, for brevity and clarity, were omitted in the main text.

\section{The model \label{sec:The-model}}

\subsection{Formulation}

Previous modeling efforts have focused on the life-cycle of the \emph{S.
titanus} with the goal of optimizing the timing of pest management
operations \cite{Maggi 2013}. Models of this kind encompass a time
frame of less than one year and are unable to describe the long-term
evolution of an infested vineyard. In spite of the complicated life
cycle, the year-to-year population levels of \emph{S. titanus} in
a vine growing area remains roughly constant, or at least of the same
degree of magnitude, if all known relevant factors (e.g. timing, number
and effectiveness of insecticidal sprays, the presence of nearby hotbeds
of infestation) are kept constant. \cite{Lessio 2011a,Lessio 2011b,Maggi 2013}.
Therefore, for time scales longer than one year, it seems to be reasonable
to formulate a continuous-time model whose variables are representative
of plants densities in a vineyard, and where the insect vector does
not appear explicitly, but is parameterized by a coupling term between
the infected and the healthy plants. 

Our model splits the grapevine population of a vineyard in four compartments
(or stages), as shown in Figure 
\begin{figure}
\begin{centering}
\includegraphics[width=0.99\columnwidth]{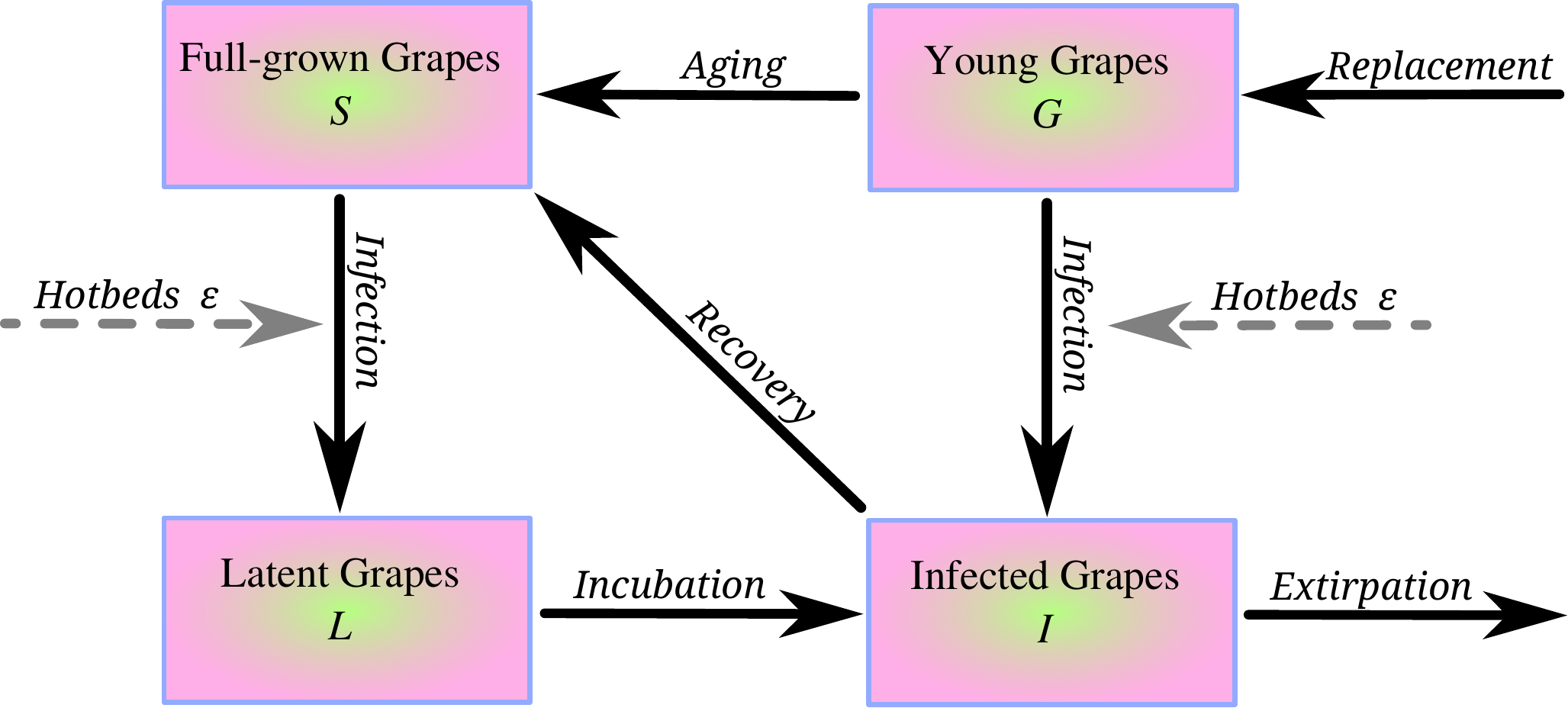}
\par\end{centering}

\caption{\label{fig:Schema}Graphical representation of the model. The grapevine
population in a vineyard is split into four compartments, representing
healthy, full-grown plants ($S$), healthy, young plants ($G$), latent
plants ($L$) and infected plants ($I$). The solid black arrows represent
the processes that increase (inward arrow) or decrease (outward arrow)
the population of each compartment. The dashed gray arrows represent
the effect of infection hotbeds close to the vineyard, triggering
the infection fluxes in initially healthy vineyards.}
\end{figure}
\ref{fig:Schema}. The variable $S$ represents the density of healthy,
full-grown plants (number of vines per unit area), and $I$ represents
the density of infected plants. Because of the lack of transovarial
transmission \cite{Alma 1997}, individuals of \emph{S. titanus} become
vectors of the phy\-to\-plasma by feeding on infected\emph{ }plants.
This occurs at the nymph stage, when the insect lacks the ability
to move from plant to plant \cite{Chuche14}. Therefore, after the
eclosion, the abundance of mobile, phy\-to\-plasma-car\-rying adults
will be proportional to the number of infected plants. Thus infection
rate of the healthy plants should be modeled by a term of the form
\begin{equation}
\mathrm{Infection\,\, rate}=f(I)S\label{eq:infection_rate}
\end{equation}
where $f$ is an unknown function that quantifies the efficiency of
phy\-to\-plasma-car\-rying adults at infecting healthy plants.
Obviously, $f$ must be a monotonically growing function of the density
of infected plants, with $f(0)=0$. Laboratory experiments show that
a small, but non-negligible fraction of plants remains healthy, after
a relatively long exposure to a population fully composed by infected
insects \cite{Schvester69}. This suggests that many probes from infected
adults may be required for a plant to eventually contract FD. If this
hypothesis is correct, then small numbers of infected plants in a
vineyard should not be very effective at spreading the disease, because
the small number of phy\-to\-plasma-car\-rying adults originating
from those plants would feed on many different healthy plants during
their lifespan, and only rarely return on the same plant enough times
to infect it. Thus we argue that also the derivative of $f$ vanishes
for $I\to0$. Of course, if the density of infected plants is large,
the probability of recurrent feeding on the same healthy plant of
phy\-to\-plasma-car\-rying insects must be large as well. Thus
we argue that $f$ should grow faster than linearly with $I$, at
least at moderately low values of $I$. The simplest mathematical
expression that captures these assumptions is
\begin{equation}
\mathrm{Infection\,\, rate}=qSI^{2}\label{eq:our_infection_rate}
\end{equation}
where $q$ is a constant whose value depends on the susceptibility
to the infection of the particular cultivar which is being considered,
and on the local abundance of \emph{S. titanus.} The value of this
constant is subject to large uncertainties. We estimate $q\approx10^{-6}$
$\mathrm{ha^{2}}$ $\mathrm{plants^{-2}}$ $\mathrm{Y^{-1}}$ , but
reasonable values range from $10^{-7}$ to $10^{-5}$ $\mathrm{ha^{2}}$
$\mathrm{plants^{-2}}$ $\mathrm{Y^{-1}}$ (see Appendix \ref{sub:An-estimate-of-q}
for details). The effect of insecticide treatments is that of lowering
the value of $q$. This is discussed in Section \ref{sub:insecticides}.

\begin{sidewaystable*}
\caption{\label{tab:TabellaParametri}Value (or range of likely values) for
the parameters appearing in the model (\ref{eq:the_model}).}

\centering{}%
\begin{tabular*}{1\textwidth}{@{\extracolsep{\fill}}rccc}
\toprule 
Process & Parameter & Value & Reference\tabularnewline
\midrule
Farmer's intervention time  & $\tau$ & $1\,\mathrm{Y}$ & %
\begin{minipage}[c]{0.35\textwidth}%
In Italy immediate eradication of infected plants is mandatory by
law (DM 32442/2000). Similar measures are in place in France.%
\end{minipage}\tabularnewline
Vineyard's design density & $D$ & $2000\,\mathrm{to}\,11000\,\mathrm{plants/ha}$ & \tabularnewline
Coupling infected-healthy & $q$ & $10^{-6}\,(10^{-7}\,\mathrm{to}\,10^{-5})\,\mathrm{ha^{2}plants{}^{-2}\, Y^{-1}}$ & \tabularnewline
Recovery from infection & $k_{1}$ & ${\displaystyle 0.4\,\mathrm{Y}^{-1}}$ & \cite{Zorloni 2008}\tabularnewline
Latency before symptoms & $k_{2}$ & ${\displaystyle \frac{1}{3}\,\mathrm{Y}^{-1}}$ & \cite{Osler 2002,Morone 2007}\tabularnewline
Mortality of infected plants & $k_{3}$ & %
\begin{minipage}[c]{0.25\textwidth}%
$\begin{cases}
\tau^{-1} & \mathrm{(managed\; vineyards)}\\
0.1\,\,\mathrm{Y}^{-1} & \mathrm{(unmanaged\; vin.)}
\end{cases}$%
\end{minipage} & %
\begin{minipage}[c]{0.35\textwidth}%
By law (DM 32442/2000)

Our estimate%
\end{minipage}\tabularnewline
Aging of new plants & $k_{4}$ & ${\displaystyle \frac{1}{5}\,\mathrm{Y}^{-1}}$ & \cite{Morone 2007}\tabularnewline
\bottomrule
\end{tabular*}
\end{sidewaystable*}

Because direct laboratory measurements of $f$ are presently lacking,
we have resisted the temptation of using more complicated functional
forms that, for example, let $f$ taper off (and maybe approach a
horizontal asymptote) as $I$ increases. At the end of Section \ref{sec:Discussion}
we discuss the effect of choosing $f$ proportional to $I$.

In the presence of hot\-beds of infection nearby the modeled vineyard
(such as infected wild grapes or an abandoned infected vineyard),
the density of infected plants that enters in the infection rate terms
should not be $I$, but rather $(I+\varepsilon)$, where the parameter
$\varepsilon$ quantifies the phy\-to\-plasma-car\-rying insects
coming from the hot\-beds, which appears to decay exponentially with
the distance of the hotbed \cite{Lessio 2014}. 

In time, some infected plants have a chance to recover from the disease,
and to return symptom-free. Furthermore, they may be re-infected,
thus recovered plants do not require a separate compartment. The process
of recovery is modeled by a flux from the $I$ to the $S$ compartments
quantified as $k_{1}I$. Experimental data, taken in vineyards where
insecticide treatments had brought to a negligible amount the presence
of \emph{S. titanus}, show that the constant $k_{1}^{-1}$ ranges
between $2$ and $3$ years for the popular \emph{Barbera} and \emph{Sauvignon}
cultivars \cite{Zorloni 2008}. For other cultivars these figures
should be taken as representative of the order of magnitude, and not
as accurate estimates of the recovery rate.

In full-grown plants, the symptoms of FD do not usually appear immediately
after the inoculation. Inoculated individuals may remain in a latent,
symptomless state for up to a few years. In our model the density
of latent plants is quantified by the compartment $L$. The amount
of latent plants that develops symptoms is quantified by the flux
$k_{2}L$ from the $L$ to the $I$ compartments. The time scale $k_{2}^{-1}$
of the process is estimated to be approximately $3$ years \cite{Osler 2002,Morone 2007}.

We assume that the farmer extirpates actively the infected plants,
on a time scale $k_{3}^{-1}=\tau$. This causes a mortality of the
infected plants quantified as $-k_{3}I$. On the same time scale,
the manager attempts to maintain a constant density $D$ of plants
in the vineyard, by planting healthy, young plants, whose density
is quantified by the variable $G$. This process continues as long
as the actual density of the vineyard (which is $S+L+I+G$) doesn't
match the desired density $D$. The constant $\tau$ quantifies the
reaction time of the farmer. While $\tau$ can't be smaller than one
year (infected grapes are roughed at the end of summer, and nursery
grapes are usually planted in the next spring in order to be productive
in autumn) it may occasionally be larger, when economic constraints
force a delay of the extirpation and replacement procedures. Young
plants are subject to infection just in the same way as full-grown
ones, with the only difference that they do not have a phase of latency,
but develop the symptoms rapidly after acquiring the phytoplasma \cite{Morone 2007}.
Thus, the process of infection produces a flux from $G$ to $I$ (rather
than to $L$). The infection rate of young plants is quantified as
$qG(I+\varepsilon)^{2}$, analogously to (\ref{eq:our_infection_rate}).
In principle we could model a different susceptibility to the infection
for the young and the full-grown plants by using different values
of the constant $q$ for the two compartments. However, lacking a
direct empirical evidence of an evident disparity in susceptibility
between young and full-grown plants, for simplicity, we prefer to
use the same value of $q$ for both. Young plants that do not become
infected eventually turn into full-grown plants by aging. This process
is modeled as a flux from the $G$ to the $S$ compartment quantified
as $k_{4}G$. For most cultivars the aging time is about $k_{4}^{-1}\approx5$
years \cite{Morone 2007}. 

The model as described by the above considerations is embodied by
the following system of first-order ordinary differential equations:

\begin{equation}
\begin{cases}
S^{\prime}= & -qS\left(I+\varepsilon\right)^{2}+k_{1}I+k_{4}G\\
L^{\prime}= & \hphantom{-}qS\left(I+\varepsilon\right)^{2}-k_{2}L\\
I^{\prime}= & \hphantom{-}qG\left(I+\varepsilon\right)^{2}+k_{2}L-k_{1}I-k_{3}I\\
G^{\prime}= & -qG\left(I+\varepsilon\right)^{2}-k_{4}G+\\
 & \hphantom{-}\tau^{-1}\left(D-S-L-I-G\right)
\end{cases}\label{eq:the_model}
\end{equation}
where the dot denotes differentiation with respect to time.

In Table \ref{tab:TabellaParametri} we summarize the above best-guesses
of the values (or value range) of the constants, deduced from evidence
given in the accompanying references.

\begin{figure*}
\begin{centering}
\includegraphics[width=0.45\textwidth]{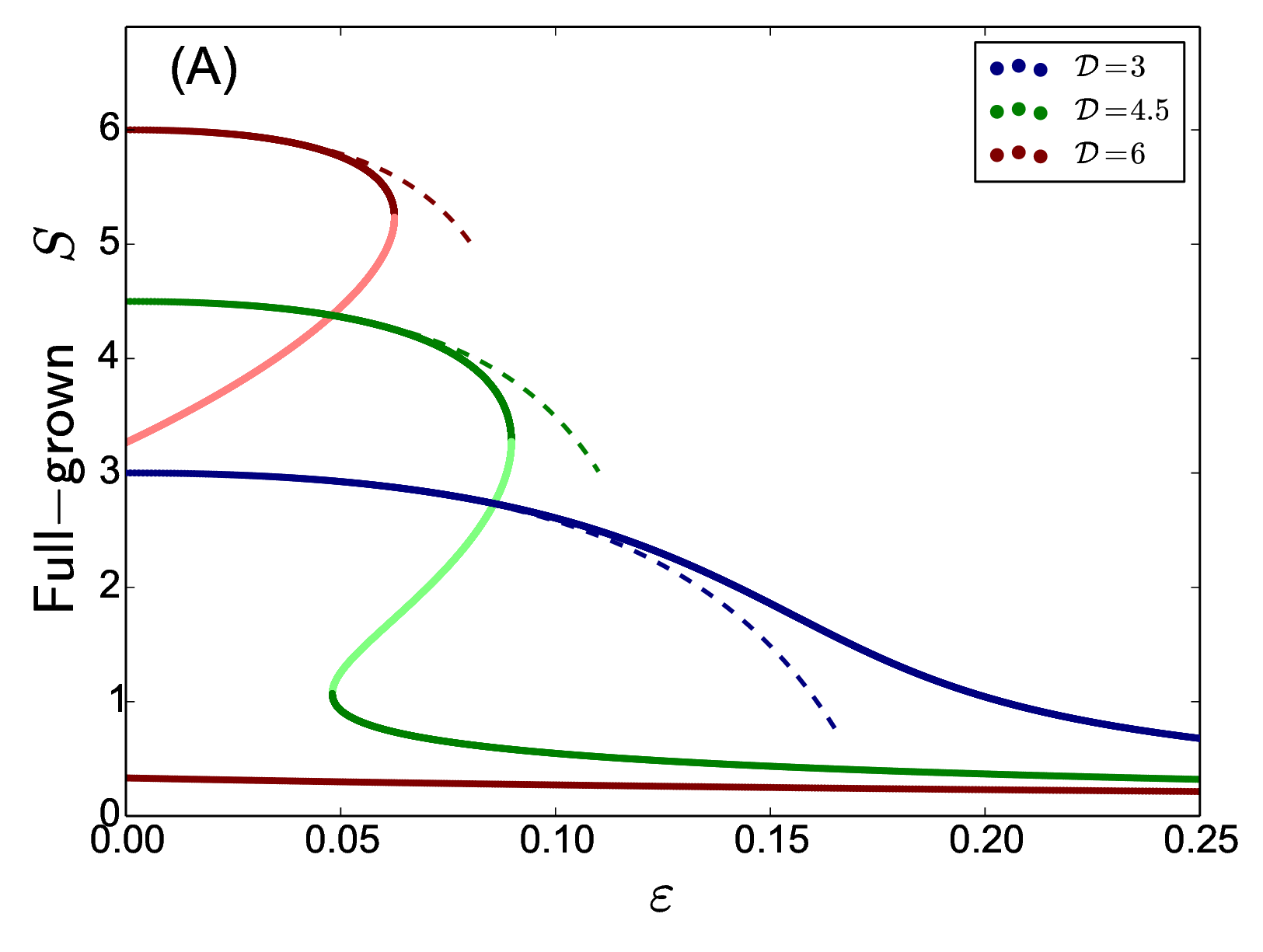}\includegraphics[width=0.45\textwidth]{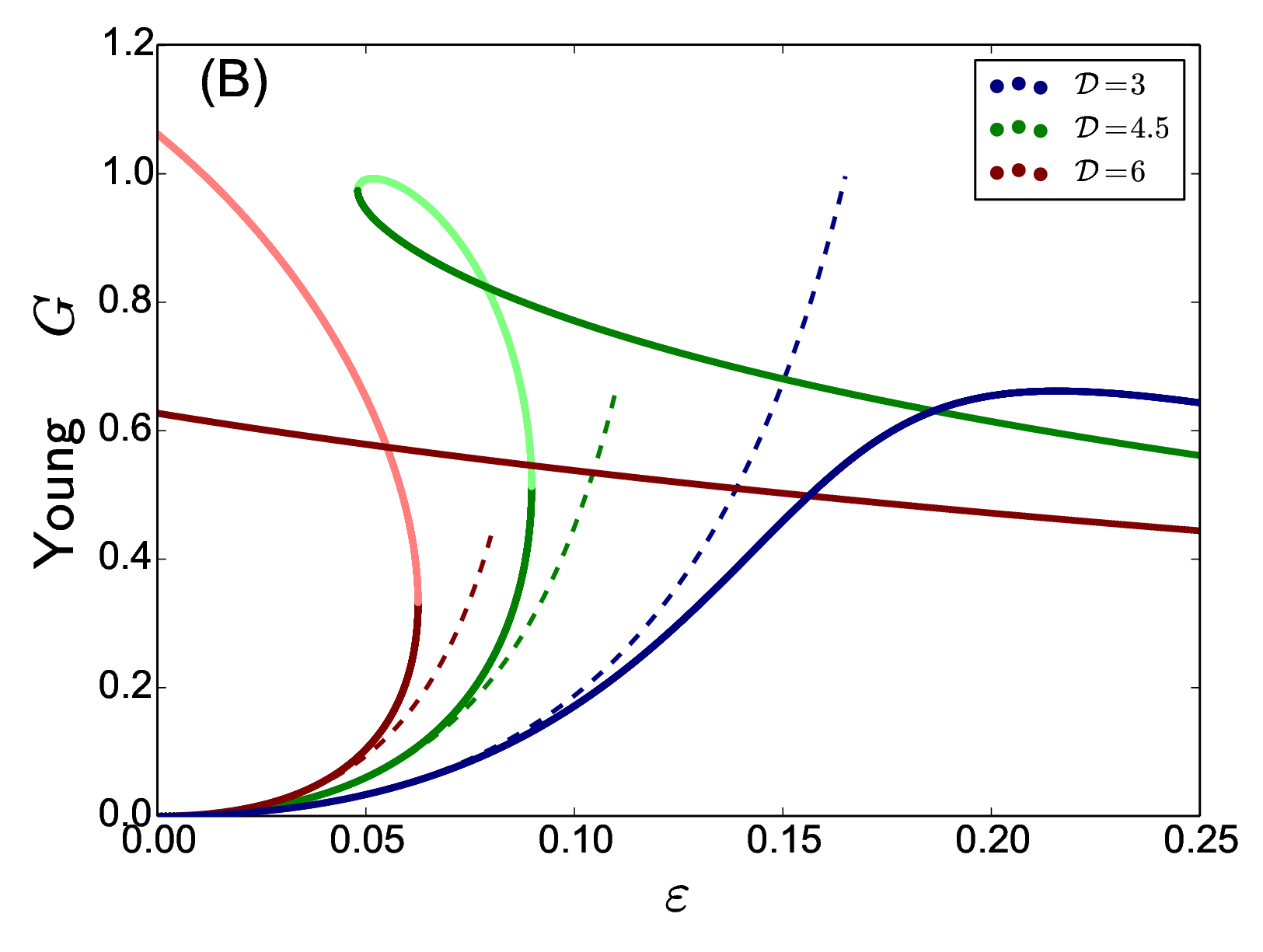}
\par\end{centering}

\begin{centering}
\includegraphics[width=0.45\textwidth]{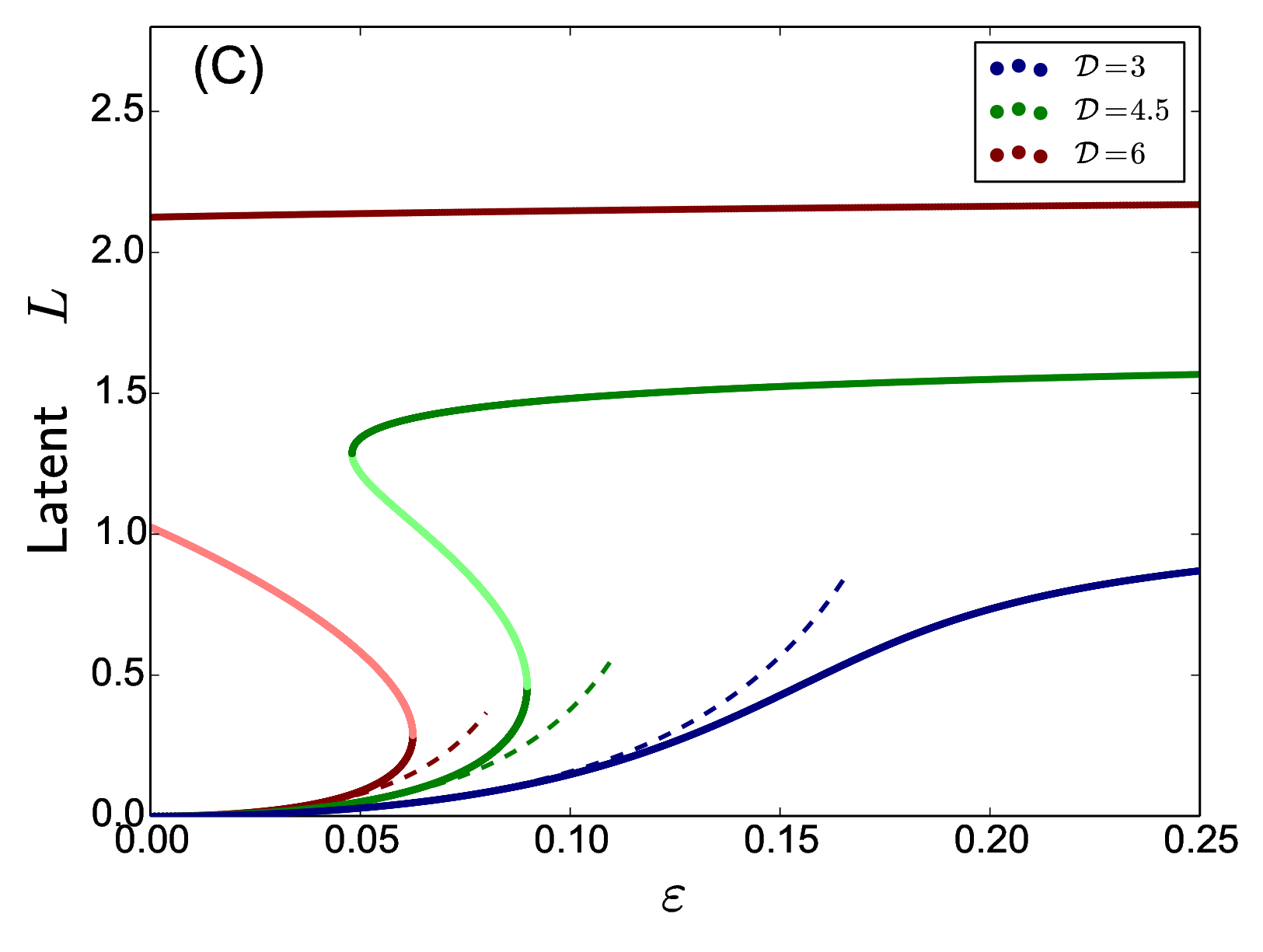}\includegraphics[width=0.45\textwidth]{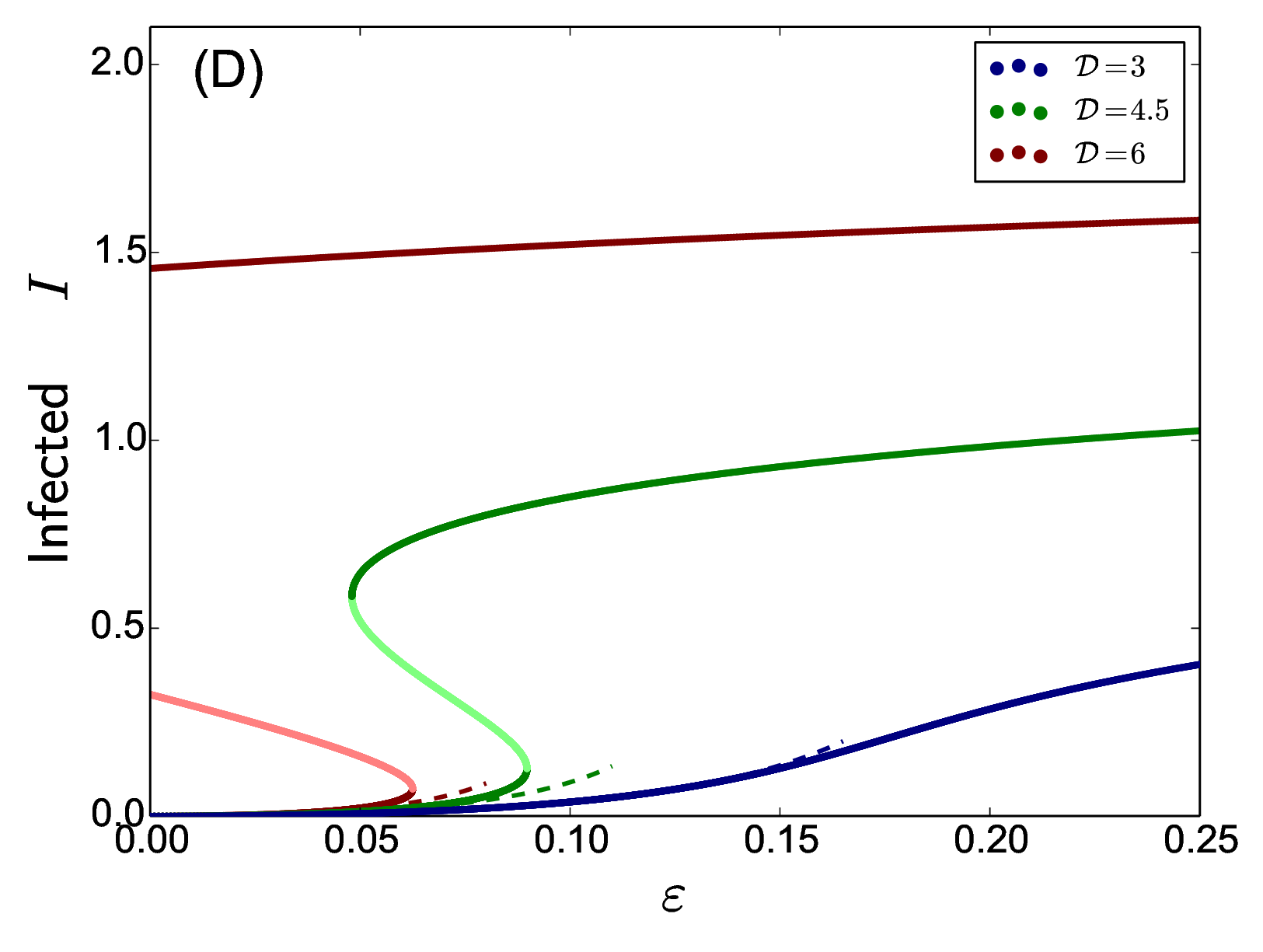}
\par\end{centering}

\caption{\label{fig:Bifurcations}Equilibria of the vineyard model (\ref{eq:nondimensional_model})
as a function of $\epsilon$ for three different values of the vineyard
density $\mathcal{D}$. Panels (A), (B), (C), (D) represent, respectively,
the density of healty, full-grown plants (the $S$ compartment); healthy,
young plants (the $G$ compartment); latent plants (the $L$ compartment);
infected plants (the $I$ compartment). Dark/light color shades represent,
respectively, stable and unstable equilibria. The dashed lines are
the approximate expressions (\ref{eq:Pad=0000E9}) of the equilibria
close to the state (\ref{eq:healthy vineyard}) of healthy vineyard.}
\end{figure*}

The system (\ref{eq:the_model}) may be brought to non-dimensional
form by using $\tau$ as the scale of time and $(q\tau)^{-1/2}$ as
the scale of grapevine density. Defining the non-dimensional quantities
\[
(\tilde{S},\tilde{L},\tilde{I},\tilde{G})=(q\tau)^{1/2}(S,L,I,G),
\]
the system in non-dimensional form reads:
\begin{equation}
\begin{cases}
\dot{\tilde{S}}= & -\tilde{S}\left(\tilde{I}+\epsilon\right)^{2}+c_{1}\tilde{I}+c_{4}\tilde{G}\\
\dot{\tilde{L}}= & \hphantom{-}\tilde{S}\left(\tilde{I}+\epsilon\right)^{2}-c_{2}\tilde{L}\\
\dot{\tilde{I}}= & \hphantom{-}\tilde{G}\left(\tilde{I}+\epsilon\right)^{2}+c_{2}\tilde{L}-c_{1}\tilde{I}-c_{3}\tilde{I}\\
\dot{\tilde{G}}= & -\tilde{G}\left(\tilde{I}+\epsilon\right)^{2}-c_{4}\tilde{G}+\\
 & \hphantom{-}\mathcal{D}-\left(\tilde{S}+\tilde{L}+\tilde{I}+\tilde{G}\right)
\end{cases}\label{eq:nondimensional_model}
\end{equation}
where the dot denotes derivation with respect to the non-dimensional
time, and the (positive) constants are $c_{1}=k_{1}\tau$, $c_{2}=k_{2}\tau$,
$c_{3}=k_{3}\tau$, $c_{4}=k_{4}\tau$, $\mathcal{D}=(q\tau)^{1/2}D$,
$\epsilon=(q\tau)^{1/2}\varepsilon$. For typographical clarity for
now on we shall omit the tildes have been omitted, and all quantities
will be in non-dimensional form, unless otherwise specified.

\subsection{Equilibria and their bifurcations\label{sub:Equilibria-and-bifurcations}}

For initial data such that $S,L,I,G\ge0$ and $S+L+I+G\le\mathcal{D}$
then the solutions of (\ref{eq:nondimensional_model}) obey the bound
$0\le S,L,I,G\le\mathcal{D}$ for all positive times (see Appendix
\ref{sub:Boundedness} for a proof). If the initial condition is such
that $S+L+I+G>\mathcal{D}$, then unacceptable solutions with negative
values may develop. However, the only occurrence in which the density
of the vineyard could be higher than the desired density $\mathcal{D}$
is when a farmer decides to thin out the vineyard in order to attain
a lower desired density. Modeling this process is, of course, well
beyond the aim of equations (\ref{eq:nondimensional_model}).

For $\epsilon=0$ the system (\ref{eq:nondimensional_model}) has
the obvious equilibrium
\begin{equation}
S=\mathcal{D},\quad L=I=G=0\label{eq:healthy vineyard}
\end{equation}
corresponding to an uninfected vineyard. Imposing the right-hand side
of (\ref{eq:nondimensional_model}) to be zero, after some algebraic
manipulations, the other equilibria of the model, for $I+\epsilon\neq0$,
may be expressed as solutions of the following system of non-linear
algebraic equations: 
\begin{equation}
\left\{ \begin{array}{l}
S={\displaystyle \frac{c_{1}I}{\left(I+\epsilon\right)^{2}}+{\displaystyle \frac{c_{3}c_{4}I}{\left(I+\epsilon\right)^{2}\left(c_{4}+\left(I+\epsilon\right)^{2}\right)}}}\\
L={\displaystyle \frac{c_{1}}{c_{2}}}I+{\displaystyle \frac{c_{3}c_{4}I}{c_{2}\left(c_{4}+\left(I+\epsilon\right)^{2}\right)}}\\
G={\displaystyle \frac{c_{3}I}{c_{4}+\left(I+\epsilon\right)^{2}}}\\
I=c_{3}^{-1}\left(\mathcal{D}-\left(S+L+I+G\right)\right)
\end{array}.\right.\label{eq:fixed_points}
\end{equation}
By substitution the first three equations of (\ref{eq:fixed_points})
into the fourth, one finds that the admissible equilibria (namely,
those that do not have negative values in any of the four compartments)
are determined by the non-negative roots of a fifth-order polynomial
in the variable $I$. For $\epsilon=0$ one of the roots is $I=0$,
which leads to the healthy vineyard equilibrium (\ref{eq:healthy vineyard}).
For small values of $\epsilon$ there exists only one equilibrium
at sufficiently small densities $\mathcal{D}$, and up to three at
higher densities. No root exists with $I\ge\mathcal{D}$. Appendix
\ref{sub:Determination-of-equilibria} gives more details on these
statements. The smallest positive root gives, for small non-zero $\epsilon$,
an equilibrium very close to (\ref{eq:healthy vineyard}). A perturbative
analysis confirms that very weak (or very far) hotbeds have almost
no effect: if $\epsilon\ll1$, then there exists an equilibrium which
differs only by $O(\epsilon^{2})$ from the healthy vineyard. Explicit,
approximate expressions of this equilibrium are given by eq. (\ref{eq:Pad=0000E9})
in Appendix \ref{sub:Pade_equilibria}.

Families of equilibria depending on a parameter may be computed numerically
\cite[Chapter 10]{Kuznetsov 1995}. Using $\epsilon$ as the control
parameter, the equilibria of the model (\ref{eq:nondimensional_model})
are shown in Figure \ref{fig:Bifurcations}, for several values of
the vineyard's desired density $\mathcal{D}$, and with $c_{1}=0.4$,
$c_{2}=1/3$, $c_{3}=1$, $c_{4}=1/5$. 

\begin{figure}
\begin{centering}
\includegraphics[width=1\columnwidth]{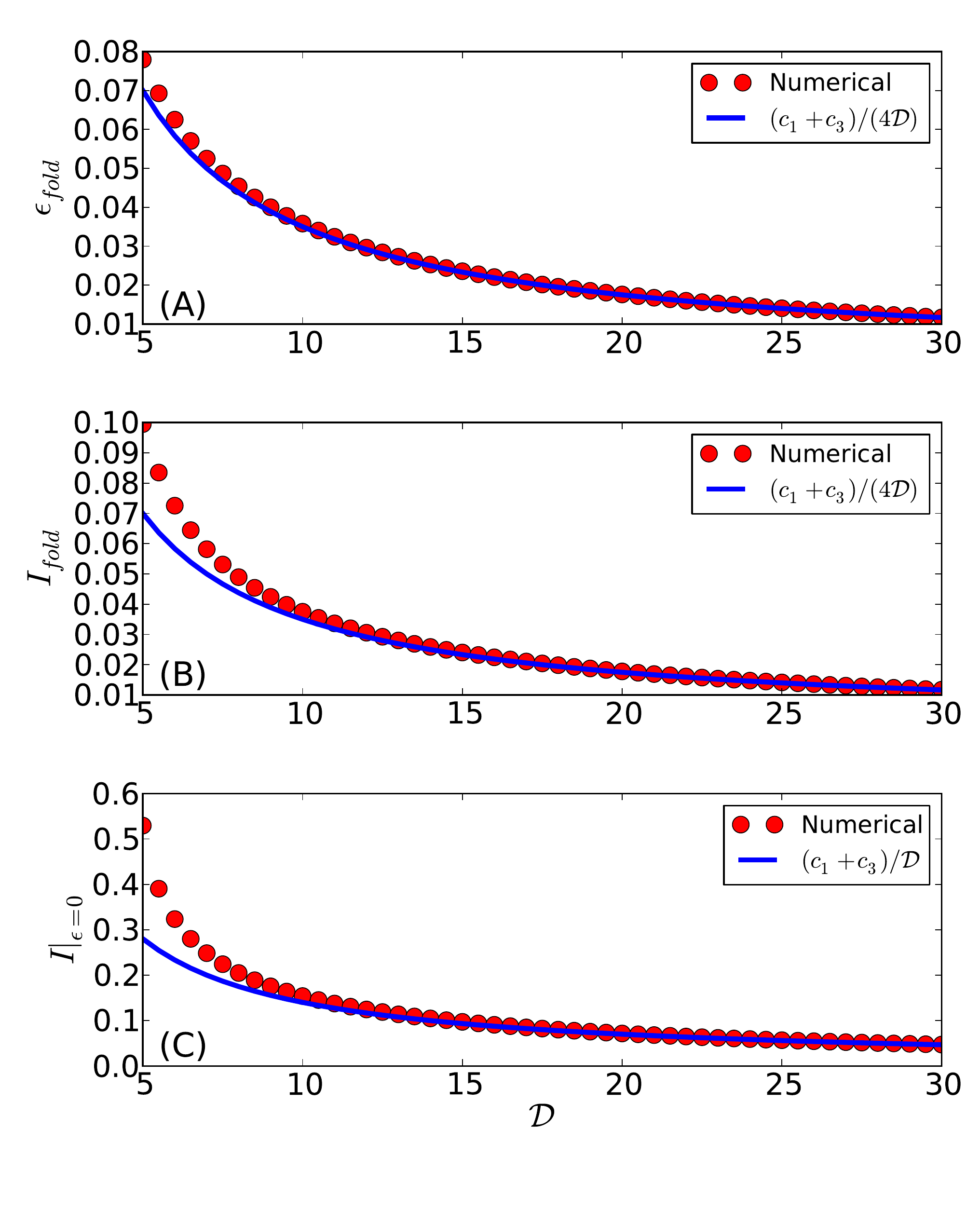}
\par\end{centering}

\caption{\label{fig:fold_approximations}Panels (A) and (B): values of $\epsilon$
and of $I$ at the saddle-node bifurcation between the healthy vineyard
branch and the saddle branch of equilibria for several values of $\mathcal{D}$.
Panel (C): value of $I$ at the saddle equilibrium with $\epsilon=0$.
The red dots are values computed numerically, the blue lines are the
approximations discussed in the Appendix \ref{sub:Approx_saddle_node}.}
\end{figure}

The linear stability of these equilibria is assessed by computing
the eigenvalues of the linearization of equations (\ref{eq:nondimensional_model})
evaluated at the equilibrium. The healthy vineyard equilibrium (\ref{eq:healthy vineyard})
has four negative real eigenvalues $\lambda_{1,2,3,4}=(-1,-c_{1}-c_{3},-c_{2},-c_{4})$,
and is therefore a stable node. For the other equilibria, shown in
Figure~\ref{fig:Bifurcations}, the eigenvalues were computed numerically.
For $\mathcal{D}\lesssim3.9$ there is only one equilibrium for each
value of $\epsilon$, which is stable. At higher densities, there
is an interval such that, if $\epsilon$ is within the interval, then
there are three equilibria (two stable nodes and a saddle); if $\epsilon$
is outside the interval, then there is only one stable equilibrium;
if $\epsilon$ is at one of the extremes of the interval, then the
saddle and one of the two nodes coalesce in a saddle-node bifurcation
(namely, the folds in Figure \ref{fig:Bifurcations} where the curves
have a vertical tangent and a stability change occurs). Only for intermediate
values of the density $\mathcal{D}$ both extremes of this interval
occur at positive values of $\epsilon$. For higher values of $\mathcal{D}$,
the branch of stable equilibria that passes through the healthy vineyard
equilibrium (\ref{eq:healthy vineyard}) folds at a positive $\epsilon$,
the other stable branch folds at a negative $\epsilon$. 

The critical values where the stable branch originating from the healthy
vineyard state (\ref{eq:healthy vineyard}) looses stability have
the following simple approximate expressions for large $\mathcal{D}$
(see Appendix \ref{sub:Approx_saddle_node} for details):
\begin{equation}
\epsilon_{fold}\approx\frac{c_{1}+c_{3}}{4\mathcal{D}},\qquad I_{fold}\approx\frac{c_{1}+c_{3}}{4\mathcal{D}}.\label{eq:approx_epsi_fold}
\end{equation}
The critical values $S_{fold}$, $L_{fold}$, $G_{fold}$ are found
by using (\ref{eq:approx_epsi_fold}) in (\ref{eq:fixed_points}).
Figure \ref{fig:fold_approximations} shows a comparison between the
critical values determined numerically and the approximation (\ref{eq:approx_epsi_fold}).

\subsection{The case of an abandoned vineyard}

Sometimes, for economic reasons, vineyards are left unmanaged. In
the absence of insecticide treatments and of active replacement of
infected plants, unmanaged vineyards may become hotbeds of infection.
A similar role is played by wild grapevines living in woodlands and
shrublands. Equations (\ref{eq:nondimensional_model}) may be used
to model these cases, simply by omitting the $G$ compartment of the
young plants. The equations then read
\begin{equation}
\begin{cases}
\dot{S}= & -SI^{2}+c_{1}I\\
\dot{L}= & \hphantom{-}SI^{2}-c_{2}L\\
\dot{I}= & -\left(c_{1}+c_{3}\right)I+c_{2}L
\end{cases}\label{eq:abandoned_vineyard}
\end{equation}
The values of the constants $c_{1}$ and $c_{2}$ may be taken the
same as before. The mortality rate of the infected plants $c_{3}$
is, instead, much smaller, because plants are not actively eradicated
once a year by a farmer, but rather die after several years of infection.
We estimate that a single full-grown plant, when infected, should
last about 10 years before dying (Table \ref{tab:TabellaParametri}).
In this very simplified approach we have omitted to introduce terms
modeling the reproduction and the natural mortality of healthy grapevines.
Owing to the long lifespan of grapevine plants, these processes occurs
on time scales which are much longer than those involving the spread
of FD, and should therefore be negligible in the present context.
For simplicity, we have also omitted any coupling term with other
nearby hotbeds: we assume that the abandoned vineyard is already infected,
and we are interested in the time evolution of the most virulent phase
of the infection, during which the abundance of infected individuals
of \emph{S. titanus} is determined by the local density of infected
plants, and any inflow from external sources becomes negligible.

\begin{figure}
\begin{centering}
\includegraphics[width=1\columnwidth]{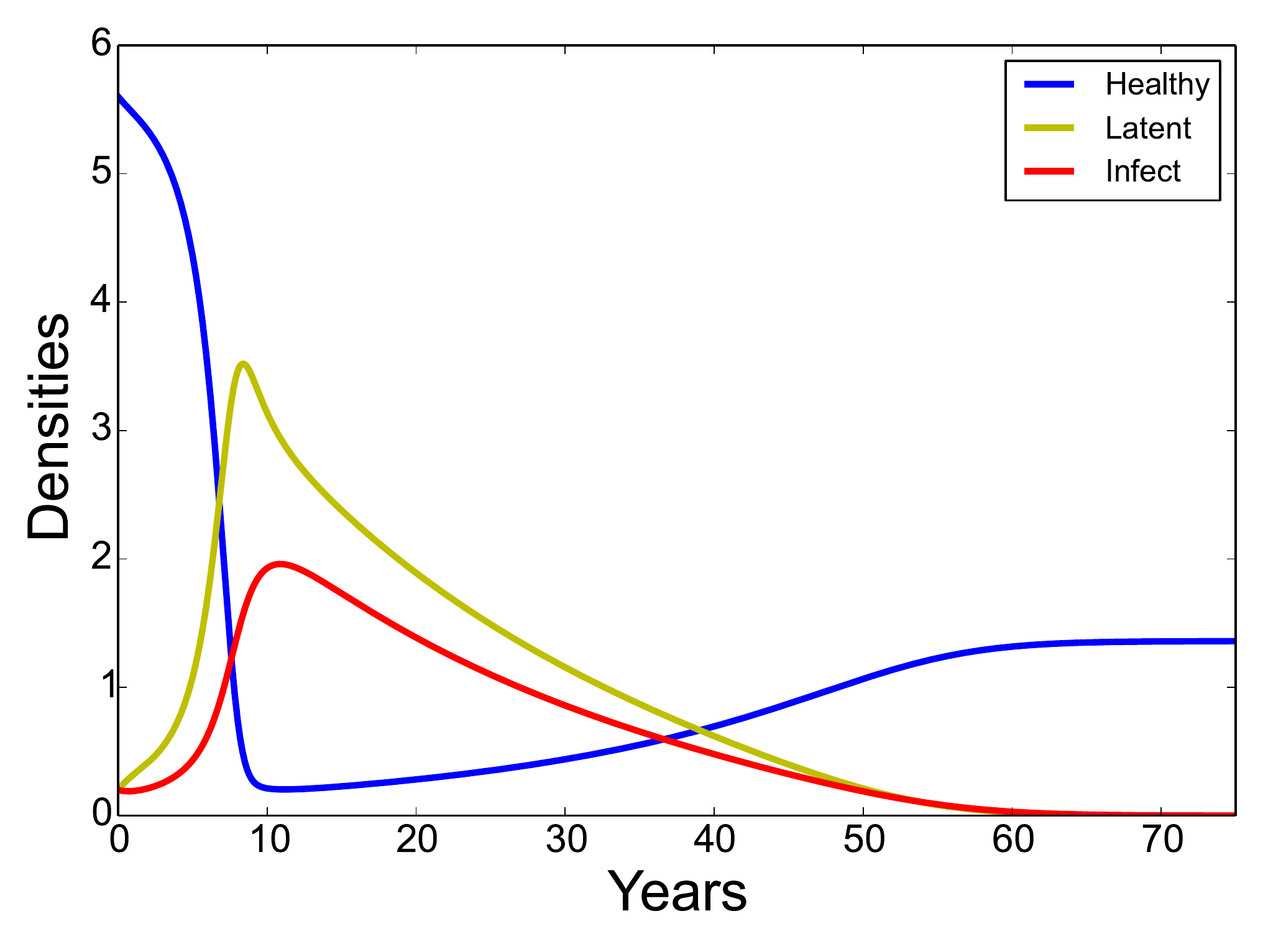}
\par\end{centering}

\caption{\label{fig:vigna_abbandonata}Numerical solution of equation (\ref{eq:abandoned_vineyard})
with the initial conditions $S(0)=5.6$, $L=0.2$, $I=0.2$. The constants
$c_{1},$ $c_{2},$ $c_{3}$ are computed from the parameter values
given in Table (\ref{tab:TabellaParametri}).}
\end{figure}

The set of states without infection $\mathcal{H}=\{S=S_{o},\, L=0,\, I=0\}$
where $S_{o}$ is an arbitrary constant, are the equilibria of the
system of equations (\ref{eq:abandoned_vineyard}). By linearizing
the equations around the equilibria we find that $\mathcal{H}$ is
a normally hyperbolic manifold having two negative eigenvalues (namely,
$\lambda_{1}=-c_{1}-c_{3}$ and $\lambda_{2}=-c2$). It is also the
center manifold of each equilibrium \cite[Chapter 5]{Kuznetsov 1995}.
Therefore, initial conditions involving a very small number of infected
and latent plants tend to fall back to an infection-free state in
$\mathcal{H}$ without experiencing an appreciable growth of infected
plants.

If the initial density of infected plants in the initial conditions
is not very small, a more complicated dynamics, illustrated in Figure
(\ref{fig:vigna_abbandonata}), will occur: from (\ref{eq:abandoned_vineyard})
we have 
\[
\frac{d}{dt}\left(L+I\right)=SI^{2}-\left(c_{1}+c_{3}\right)I,
\]
thus, if initially it is $SI>c_{1}+c_{3}$, then the density of latent
and infected plants will continue to grow as long as the latter inequality
is satisfied. This produces, in the span of a few years, a dramatic
decrease of the density of healthy plants mirrored by a corresponding
rise of infected and latent plants. As the number of infected plants
increases, so does the number of plants that recover and become healthy
again. The epidemic peaks when the recovered plants become a substantial
fraction of the healthy plants. This is in qualitative agreement with
the results of the experiments of Morone et al. \cite{Morone 2007}.
After this rapid, virulent phase, the density of healthy plants has
dropped so much that $SI<c_{1}+c_{3}$. The density of latent and
infected plants slowly decreases, whereas the density of healthy plants
experiences a slow growth, thanks to the recovery of previously infected
plants. Eventually, after an almost century-long transient, the abandoned
vineyard returns to a healthy state, but with a drastically lower
plant density.

\section{Discussion\label{sec:Discussion}}

\subsection{Practical implications of the structure of the bifurcation diagram}

The bifurcation analysis of Section \ref{sub:Equilibria-and-bifurcations}
shows the crucial importance of infection hotbeds in determining the
state of a nearby vineyard. If the hotbeds are absent or weak (that
is, if the value of $\epsilon$ is small), then there is always a
stable state which doesn't differ much from the healthy vineyard state:
infected plants are very few and replacing them with young ones keeps
the infection at very low levels. 

If $\epsilon$ increases, the outcome depends on the value of the
parameter $\mathcal{D}$ (the non-dimensional design density of the
vineyard). For densities $\mathcal{D}\lesssim3.9$ (which, according
to our estimate of $q$, correspond to $D\lesssim3900\;\mathrm{plants\, ha^{-1}}$)
the number of healthy plants decreases gradually with increasing $\epsilon$:
one would observe a progressive worsening of the vineyard's health
as the hotbeds become stronger (see the blue curve in Figure \ref{fig:Bifurcations}).
For higher densities, the gradual decrease occurs only up to the critical
value $\epsilon_{fold}$. When the strength of the hotbeds becomes
such that $\epsilon>\epsilon_{fold}$, then the stable fixed point,
corresponding to a vineyard with just a few infected plants, disappears.
As a consequence, the vineyard undergoes a transition towards the
only stable fixed point left, corresponding to a state dominated by
infected and latent plants. 

\begin{figure*}
\begin{centering}
\includegraphics[width=1\textwidth]{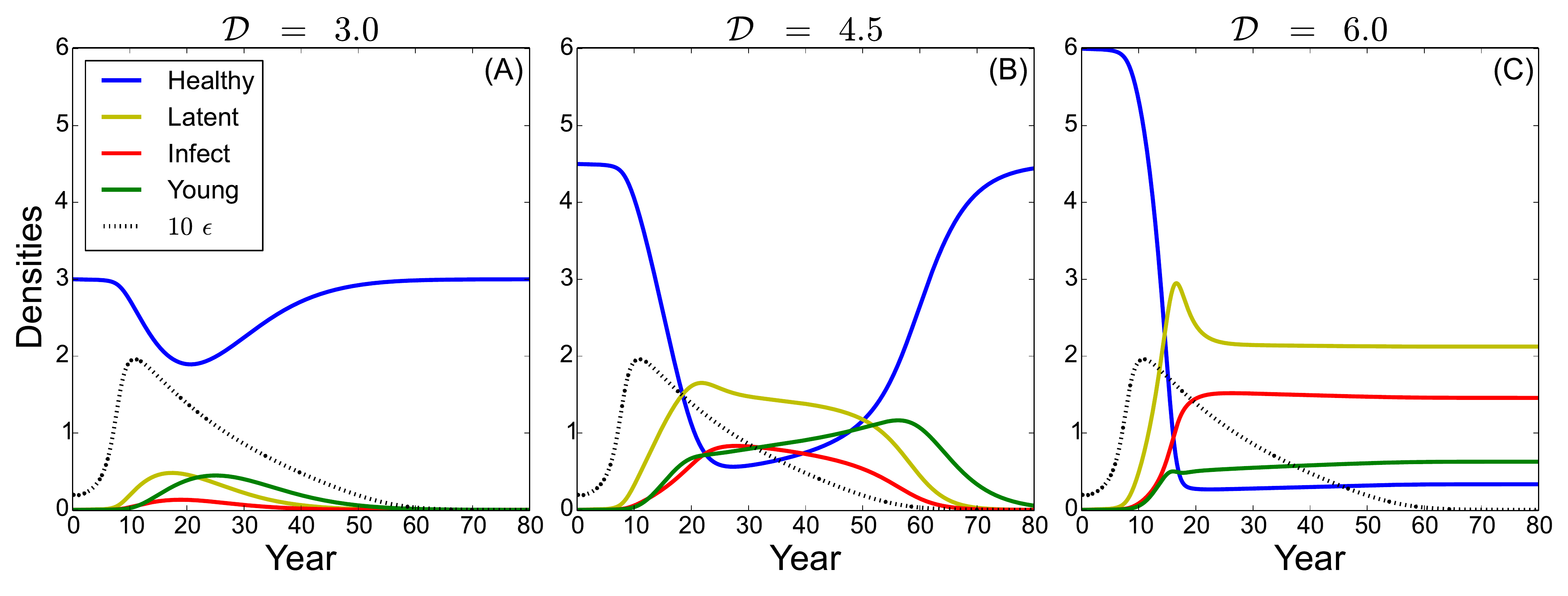}
\par\end{centering}

\caption{\label{fig:Time_evolution}Time evolution of of a vineyard with design
density $\mathcal{D}=3$ (A), $\mathcal{D}=4.5$ (B), and $D=6$ (C)
exposed to hotbeds with a time-varying strength $\epsilon$. Note
that, for clarity, the value of $\epsilon$ (represented by the dotted
black line) is multiplied by ten.}
\end{figure*}

After the transition, if the strength of the hotbeds eventually decreases
and returns to zero, the outcome also depends on the design density
of the vineyard. If $\mathcal{D}\lesssim4.8$ (corresponding to $D\lesssim4800\;\mathrm{plants\, ha^{-1}}$)
then there exists another critical value of $\epsilon$, such that,
when the strength of the hotbeds is lower than this critical value,
the system experiences a transition back to the state with just a
few infected plants (see the green curves in Figure \ref{fig:Bifurcations}).
In this case, one would observe a classic hysteresis cycle. For higher
densities (see the red curves in Figure \ref{fig:Bifurcations}) performing
a complete cycle is impossible: the second critical value of $\epsilon$
is negative, which is impossible to attain, because it would correspond
to a meaningless negative density of infected plants in the hotbeds.
Therefore, even if all the hotbeds were removed (that is, $\epsilon=0$)
the vineyard would remain stuck in the stable state with very few
healthy plants. 

In order to further elucidate this dynamics, and to show how the structure
of the bifurcation diagram shapes the time evolution of a vineyard
exposed to a nearby hotbed, we have solved numerically the equations
(\ref{eq:nondimensional_model}) with a time varying $\epsilon$.
We have set $\epsilon=c_{o}I_{a}$, where $I_{a}$ is the density
of infected plants in an abandoned vineyard according to the model
(\ref{eq:abandoned_vineyard}), with the parameters and initial conditions
of Figure \ref{fig:vigna_abbandonata}. The strength of the hotbed
peaks about ten years after the beginning of the simulation, then
slowly returns to zero. By setting $c_{o}=0.1$, and keeping all the
other parameters of the managed vineyard as in Figure \ref{fig:Bifurcations},
we obtain the results shown in Figure \ref{fig:Time_evolution}. 

At low densities (Figure \ref{fig:Time_evolution} A) there is only
one stable fixed point for each value of $\epsilon$. Thus, as $\epsilon$
varies in time, the state of the vineyard changes gradually, with
only a moderate loss of healthy plants. When $\epsilon$ returns to
zero, the vineyard returns to a healthy state.

At intermediate densities (Figure \ref{fig:Time_evolution} B), after
the initial transient during which the number of healthy plants decreases
very slowly, the strength of the hotbed crosses the bifurcation point.
The vineyard then experiences a rapid transition to the state with
very few healthy plants. When the strength of the hotbed decreases
below the second critical value, then the vineyard slowly recovers,
and returns to a healthy state, closing the hysteresis cycle. 

At high densities (Figure \ref{fig:Time_evolution} C), there is also
a rapid transient to the state with few healthy plants, but then the
vineyard never recovers the healthy state, even for vanishingly low
strengths of the hotbed, because the second critical point is unreachable
and the hysteresis loop can not be closed.

\subsection{The effect of the insecticides\label{sub:insecticides}}

Because insecticides cause a drop in the number of adults of \emph{S.
titanus}, they diminish the strength of the coupling between the infected
and the healthy plants. In our model this translates in a decrease
of the value of the parameter $q$ which appears in equations (\ref{eq:the_model}). 

The consequences of this decrease are best understood by recalling
that $q$ appears in the scale of density used to define the non-dimensional
compartments. In particular, the non-dimensional design density and
the dimensional one, are linked by the following identity
\begin{equation}
\mathcal{D}=\left(q\tau\right)^{1/2}D.\label{eq:non-dimensional_D}
\end{equation}
It follows that a decrease in $q$ is equivalent to a decrease in
the \emph{non-dimensional} value of the vineyard's design density
$\mathcal{D}$. Thus, the persistent use of insecticides may have
the effect of changing the shape of the bifurcation diagram of a given
vineyard from, say, one that looks like the red curves in Figure (\ref{fig:Bifurcations})
to one that looks like the blue curves. 

In this respect, insecticides should be seen as beneficial, because,
by potentially removing the fold bifurcations, they avoid catastrophic
transitions from a state in which the vineyard is close to the healthy
state, to one in which most of the plants are either infected or latent.
However, the insecticide treatment can not be interrupted while there
are still nearby hotbeds of infection, because the bifurcation diagram
would spring back to its original shape, and the state of the vineyard
would rapidly deteriorate, even if it had attained an almost perfect
recovery during the years of insecticide treatment.

On the other hand, if the hotbeds of infection were gradually eliminated,
any vineyard still in relatively good health would return to the healthy
state, without the need of insecticides, thanks just to the continued
action of extirpation of the infected plants and replacement with
the young ones. Therefore, the elimination of hotbeds as a source
of inoculum seems to be crucial in maintaining a healthy status in
vineyards.

In the situations in which a decrease of the value of $\mathcal{D}$
is desirable, this can also be achieved by decreasing the design density
$D$ of the vineyard. In order to do so, a farmer may replace with
young plants only a fraction of those eradicated because infected.
This procedure has the added benefit of reducing the number of young
plants in the vineyard, which, being not subject to a latency period,
may immediately become infected, thus helping spreading the disease.
Obviously, the decision of decreasing the design density of a vineyard
has to be evaluated also on the basis of economic considerations,
but it seems unwise to unconditionally exclude this option, and rely
exclusively on insecticide treatments.

\subsection{The functional form of the infection rate}

We have argued that the function $f$ appearing in the infection rate
(\ref{eq:infection_rate}) should grow faster than linearly with $I$
at low numbers of infected plants. In order to support our argument
we have also investigated the case in which the infection rate is
\begin{equation}
\mathrm{Infection\,\, rate}=\hat{q}SI\label{eq:wrong_infection_rate}
\end{equation}
where $\hat{q}$ is expressed as $\mathrm{ha\, plants^{-1}Y^{-1}}$.
Thus a scale of grapevine density is $(\hat{q}\tau)^{-1}$. Using
(\ref{eq:wrong_infection_rate}) in place of (\ref{eq:our_infection_rate}),
the non-dimensional model (\ref{eq:nondimensional_model}) becomes
the following
\begin{equation}
\begin{cases}
\dot{S}= & -S\left(I+\hat{\epsilon}\right)+c_{1}I+c_{4}G\\
\dot{L}= & \hphantom{-}S\left(I+\hat{\epsilon}\right)-c_{2}L\\
\dot{I}= & \hphantom{-}G\left(I+\hat{\epsilon}\right)+c_{2}L-c_{1}I-c_{3}I\\
\dot{G}= & -G\left(I+\hat{\epsilon}\right)-c_{4}G+\\
 & \hphantom{-}\hat{\mathcal{D}}-\left(S+L+I+G\right)
\end{cases}\label{eq:nondimensional_model-propto-I}
\end{equation}
where $\left(\hat{\mathcal{D}},\hat{\epsilon}\right)=\hat{q}\tau\left(D,\varepsilon\right)$.
The state of healthy vineyard (\ref{eq:healthy vineyard}) is obviously
an equilibrium also for equations (\ref{eq:nondimensional_model-propto-I}).
A linear stability analysis shows that the linearization around this
equilibrium has three negative eigenvalues. The fourth eigenvalue
is
\[
\lambda_{4}=\frac{\sqrt{4c_{2}\left(\mathcal{D}-c_{1}-c_{3}\right)+\left(c_{1}+c_{2}+c_{3}\right)^{2}}-\left(c_{1}+c_{2}+c_{3}\right)}{2}.
\]
The healthy vineyard is the unstable if $\lambda_{4}>0$, that is
if
\begin{equation}
\mathcal{D}>c_{1}+c_{3}.\label{eq:instability_criterion}
\end{equation}
A bifurcation analysis analogous to that of Section \ref{sub:Equilibria-and-bifurcations}
shows that the state of healthy vineyard is part of a family of equilibria
which exists for any $\epsilon\ge0$ and is not subject to any bifurcation. 

When (\ref{eq:instability_criterion}) is satisfied, this family of
equilibria is unstable. In addition, the $G$ compartment assumes
negative values for $\epsilon>0$, which makes these equilibria ecologically
meaningless. For the same parameters, there is a second family of
equilibria which is stable, ecologically acceptable, and corresponds
to a vineyard dominated by infected and latent plants.

If (\ref{eq:instability_criterion}) is not satisfied, and $\mathcal{D}<c_{1}+c_{3}$
holds, then the family of equilibria containing the healthy vineyard
is stable, and no compartment assumes negative values for any value
of $\epsilon$. For increasing $\epsilon$, the density of healthy
plants gradually decreases and that of latent and infected plants
gradually increases.

An analysis analogous to that reported in Section \ref{sub:An-estimate-of-q}
leads to an estimate of $\hat{q}$ in the range $10^{-3}$ to $10^{-2}$
$\mathrm{ha\, plants^{-1}Y^{-1}}$. Estimating $c_{1}+c_{3}\approx1.4$
(see Table \ref{tab:TabellaParametri}), this suggests that the typical
healthy vineyard should be unstable to infinitesimal perturbations.
As a consequence, even a small number of infected adults of \emph{S.
titanus} would precipitate a healthy vineyard into the stable state
dominated by infected plants, on a time scale proportional to $\lambda_{4}^{-1}$.
If, on the other hand, $\hat{q}$ were as small as to make stable
the state of healthy vineyard, then a vineyard would never become
dominated by infected plants, except when surrounded by very strong
and virulent hotbeds.

\begin{table}
\caption{\label{tab:Infection_spread}For seven Italian regions we report the
date of the first known appearence of \emph{Flavescence Dorée}, and
the date of the beginning of the epidemics. In the region Marche,
as of today, three well-established hotbeds have been reported, but
not a widespread epidemics. For Valle d'Aosta the beginning of the
epidemics is estimated from personal communications.}

\centering{}%
\begin{tabular*}{1\columnwidth}{@{\extracolsep{\fill}}rcc}
\toprule 
Region & %
\begin{minipage}[c]{8em}%
\begin{center}
First known\\
 appearence
\par\end{center}%
\end{minipage} & %
\begin{minipage}[c]{8em}%
\begin{center}
Beginning of the\\
 epidemics
\par\end{center}%
\end{minipage}\tabularnewline
\midrule
Veneto  & 1976 \cite{Egger 1983} & 1980-82 \cite{Belli 1983}\tabularnewline
Trentino  & 1985 \cite{Mescalchin 1986} & 1992 \cite{Vindimian 1983}\tabularnewline
Friuli  & uncertain  & 1986 \cite{Carraro 1986}\tabularnewline
Marche  & 2001 \cite{Credi 2002} & not yet \cite{Riolo 2014}\tabularnewline
Piemonte  & 1978 \cite{Belli 1978} & 1998 \cite{Morone 2007}\tabularnewline
Valle d'Aosta & 2006 \cite{Bonfanti 2008} & 2010 \tabularnewline
 Lombardia  & 1972 \cite{Belli 1973} & 1985 \cite{Fortusini 1988}\tabularnewline
\bottomrule
\end{tabular*}
\end{table}

These findings are at odds with the observed phenomenology of infestations
of FD. If the model (\ref{eq:nondimensional_model-propto-I}) were
correct, when exposed to FD, a well-managed vineyard would either
develop just a small number of infected plants, without further progresses
(stable healthy state), or, more likely, rapidly experience a destructive
infestation (unstable healthy state). A transition from the first
to the second situation (as shown in Figure \ref{fig:Time_evolution}
B, C) appears to be impossible if one accepts the expression (\ref{eq:wrong_infection_rate})
for the infection rate. An example of this transition is reported
for Serbia, where several vineyards of cultivar \textquotedblleft{}Plovdina\textquotedblright{},
very sensitive to FD, raised up to a 100\% symptomatic grapes in three
years starting from an infection rate lower than 5\% \cite{Kuzmanovic08}. 

Table \ref{tab:Infection_spread} shows that, in Italian wine-making
regions, there has always been a delay of several years from the date
of the first report of FD, to the date of the beginning of the epidemic
status. The latter may be defined as when FD starts to spread all
over the territory without being confined to a few hot points. Generally,
the epidemic status is recognized when a consistent part of the vineyards
has more than 30\% of infected plants, and eradication is not considered
possible anymore. For instance, in Piedmont, the territory is classified
as: free but vulnerable areas (FD absent, but very likely to occur);
hotspot areas (FD present but not settled, possible to eradicate);
settlement areas (FD settled, impossible to eradicate). 

If healthy vineyards were unstable with respect to small inflows of
infected adults of \emph{S. titanus}, one would expect a much more
rapid development of the epidemics. The observed slow progression
of the infestation, together with the fact that, in the presence of
an epidemics, the infestation in a vineyard is unlikely to remain
limited to a small portion of the plants, but generally progresses
up to a state dominated by infected plants (sometimes even if insecticides
are used), strongly points to the existence of a critical transition
between two alternative stable states of almost-healthy vineyard and
completely infested vineyard, as suggested by the bifurcation diagram
of Figure \ref{fig:Bifurcations}.

\begin{figure*}
\begin{centering}
\includegraphics[width=1\textwidth]{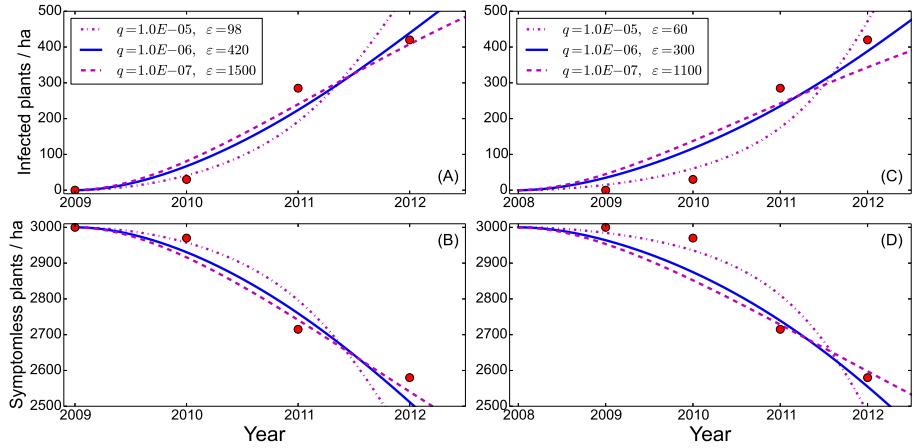}
\par\end{centering}

\caption{\label{fig:fitting_q}Comparison between observed data (dots) and
the model (lines) for different values of the parameter $q$. In panels
(A), (C) the dots represent the observed density of infected plants
in an experimental vineyard, the lines are the $I$ compartment. In
panels (B), (D) the dots represent the observed density of symptomless
plants, the lines are the sum of the $S$ and $L$ compartments. In
panels (A), (B) the numerical solution starts in 2009, and uses the
$q\;\left(\mathrm{ha^{2}plants^{-2}Y^{-1}}\right)$ and $\varepsilon\;\left(\mathrm{plants\, ha^{-1}}\right)$
values given in the inset of panel (A). In panels (C), (D) the solution
starts in 2008 and the $q,\;\varepsilon$ values are given in the
inset of panel (C).}
\end{figure*}

\section{Conclusions\label{sec:Conclusions}}

We have developed a model for the time evolution of a FD epidemics
in a vineyard. The presence of the vector of the disease (the leafhopper
\emph{Scaphoideus titanus} Ball) is not explicitly modeled, but is
parameterized as an interaction term between the infected and the
healthy grapevine plants. The presence of infection hotbeds near the
vineyard appears as a parameter in this interaction term. In addition
to infection, the model takes also into account incubation, recovery
and aging processes, and also extirpation and replacement of infected
plants, operated by the farmer. 

The model shows that, in the presence of abundant populations of \emph{S.
titanus}, or, equivalently, for vineyards with high plant density,
two stable equilibria are possible. One of these corresponds to a
situation with just a few infected plants, where the infection is
kept under control by the extirpation and replacement process. The
other equilibrium corresponds to a vineyard dominated by infected
plants, where extirpation and replacement is ineffective. When the
strength of the hotbeds crosses a critical threshold, only the latter
equilibrium survives, and the former disappears. Therefore, vineyards
infected with FD may undergo an irreversible transition from a near-healthy
state to a severely compromised one. 

The model suggests that insecticide treatments, together with continued
extirpation and replacement of infected plants, may be an effective
way to recover from a severe infestation. However, the model also
demonstrates that insecticides are just a stopgap measure. Without
the elimination of the hotbeds, FD would not disappear from an infested
vineyard, and any weakening of the treatments may precipitate the
situation again. This finding calls for a proactive search and removal
of any infection hotbeds as soon as the presence of FD becomes apparent
in a grapevine-farming territory. These preventive measures, if successful,
could avoid the need of extensive insecticide treatments on productive
vineyards.

\subsection*{Acknowledgements}
This research was conducted within the frame of the project: \textquotedbl{}Elaborazione
e stesura di un protocollo di lotta guidata alla Flavescenza dorata
e al suo vettore Scaphoideus titanus nella zona di produzione dell'Asti
docg\textquotedbl{}, funded by \textquotedbl{}Consorzio per la tutela
dell'Asti docg\textquotedbl{}. We thank Jost von Hardenberg for a
helpful discussion on this problem.

\section{Appendix\label{sec:Appendix}}

\subsection{An estimate of the value of $q$\label{sub:An-estimate-of-q}}

As part of an experiment in the province of Cuneo (Italy), a small
vineyard of $0.475$ ha was monitored from 2009 to 2012. The vineyard
had an initial density $D=3000$ plants/ha and no infected plants.
Flavescence Dorée was already well established in the surrounding
territory. However we lack quantitative data allowing for the estimation
of the appropriate value of $\varepsilon$. No insecticide treatments,
nor extirpation of infected plants was performed. Every year, the
number of infected and simptomless plants was assessed (the red dots
in Figure \ref{fig:fitting_q} (A), (C) and (B),(D), respectively).
The assessment would not distinguish between healthy and latent plants,
both classified as symptomless.

In order to determine a reasonable range of values of $q$, we apply
the model (\ref{eq:the_model}) by setting $G=0$ and dropping the
last equation (which models the replacement of infected plants with
young ones). Figure (\ref{fig:fitting_q}) shows a comparison between
the model results and the observed data, for several choices of the
parameters $q$ and $\varepsilon$. The other coefficients are those
of Table \ref{tab:TabellaParametri}. The density of symptomless plants
is compared with the sum of the $S$ and $L$ compartments. The initial
condition is $S=D$, $L=I=0$. In Figures \ref{fig:fitting_q} (A),
(B) the numerical solution starts in 2009, the last year without infected
plants. In Figures \ref{fig:fitting_q} (C), (D) the solution starts
in 2008. This allows for the hypothesis that for one year all the
inoculated plants remained in the latent state, or with symptoms as
weak as to evade detection. In the first case, $q\approx10^{-6}\,\mathrm{ha^{2}plants{}^{-2}\, Y^{-1}}$
gives a reasonable fit of the data, while in the second a value as
high as $q\approx10^{-5}\,\mathrm{ha^{2}plants{}^{-2}\, Y^{-1}}$
yields a more convincing fit. Values as low as $q\approx10^{-7}\,\mathrm{ha^{2}plants{}^{-2}\, Y^{-1}}$
also give an acceptable fit, if the initial condition refers to 2009,
but should probably be ruled out, because they require unrealistically
high values of $\varepsilon$.

\subsection{Boundedness and non-negativity of the solutions\label{sub:Boundedness}}

For non-negative initial conditions such that $S+L+I+G\le\mathcal{D}$
the solutions of the model equations (\ref{eq:nondimensional_model})
remain non-negative and bounded by $\mathcal{D}$ at all later times.
In fact, by adding together the four equations in (\ref{eq:nondimensional_model}),
and defining the total vineyard density $x=S+L+I+G$, we obtain
\begin{equation}
\dot{x}=\mathcal{D}-x-c_{3}I.\label{eq:add_all_up}
\end{equation}
Considering $I$ as a known function of time, we have that the solution
of (\ref{eq:add_all_up}) is 
\begin{equation}
x(t)=\mathcal{D}+\left(x(0)-\mathcal{D}-c_{3}\int_{0}^{t}e^{s}I(s)\,\mathrm{d}s\right)e^{-t}.\label{eq:add_up_sol}
\end{equation}
This shows that, if the initial vineyard density is $x(0)\le\mathcal{D}$,
then, as long as $I$ remains non-negative, it will be $x(t)\le\mathcal{D}$.
But if at any time $t$ we have $S,L,I,G\ge0$ and $x(t)\le\mathcal{D}$,
then from (\ref{eq:nondimensional_model}) we deduce $S=0\Rightarrow\dot{S}\ge0$,
$L=0\Rightarrow\dot{L}\ge0$, $I=0\Rightarrow\dot{I}\ge0$, and $G=0\Rightarrow\dot{G}\ge0$.
Therefore, none of the four compartments can become negative. Thus
we have that $0\le x(t)\le\mathcal{D}$ at all times, which implies
$0\le S,L,I,G\le\mathcal{D}$.

\subsection{\label{sub:Determination-of-equilibria}Determination of the equilibria
of the model}

By substituting the first three expressions of (\ref{eq:fixed_points})
in the fourth, and then multiplying by $c_{2}\left(I+\epsilon\right)^{2}\left(\left(I+\epsilon\right)^{2}+c_{4}\right)$
we obtain that the equilibrium densities of infected plants are the
non-negative roots of the fifth-order polynomial $\mathcal{P}(I)=\sum_{n=0}^{5}q_{n}I^{n}$,
whose coefficients are
\begin{equation}
\begin{array}{l}
q_{5}=\hphantom{-}c_{2}\left(c_{3}+1\right)+c_{1}\\
q_{4}=-c_{2}\mathcal{D}+4q_{5}\epsilon\\
q_{3}=\hphantom{-}c_{2}\left(c_{3}+1\right)c_{4}+\left(c_{1}+c_{3}\right)\left(c_{2}+c_{4}\right)+4q_{4}\epsilon-10q_{5}\epsilon^{2}\\
q_{2}=-c_{2}c_{4}\mathcal{D}+2q_{3}\epsilon-2q_{4}\epsilon^{2}\\
q_{1}=\hphantom{-}c_{2}\left(c_{3}+c_{1}\right)c_{4}+2q_{2}\epsilon-3q_{3}\epsilon^{2}+4q_{4}\epsilon^{3}-5q_{5}\epsilon^{4}\\
q_{0}=-c_{2}c_{4}\mathcal{D}\epsilon^{2}-c_{2}\mathcal{D}\epsilon^{4}
\end{array}\label{eq:coefficients}
\end{equation}
Recalling that $c_{1},\ldots,c_{4}>0$ and $\mathcal{D}>0$, from
the last equation in (\ref{eq:the_model}) it follows that there are
no equilibria with $I\ge\mathcal{D}$ and $S,L,G\ge0$. Because from
(\ref{eq:fixed_points}) it follows that to any non-negative root
of $\mathcal{P}$ corresponds an equilibrium with non-negative values
for all the four compartments, then we deduce that $\mathcal{P}$
cannot have real roots larger than $\mathcal{D}$.

Note that the coefficients $q_{5},\ldots,q_{0}$ are polynomials in
$\epsilon$. We observe that, for any given $\mathcal{D}$, there
are sufficiently small values of $\epsilon$ so that the coefficients
of the odd powers are positive and those of the even powers are negative.
Then, from Descartes' rule of signs, it follows that $\mathcal{P}$
has no negative roots. Hence, being an odd-degree polynomial, it must
have at least one non-negative real root. In the special case $\epsilon=0$
then $q_{0}=0$, and a real root is $I=0$, which yields the equilibrium
(\ref{eq:healthy vineyard}). We also observe that for any positive
value of $\epsilon$ as small as to make $q_{3},q_{1}>0$, $2q_{3}>q_{4}\epsilon$,
there exist sufficiently small values of $\mathcal{D}$ such that
$q_{4},q_{2}>0$. Then, from Descartes' rule of signs, it follows
that $\mathcal{P}$ has one, and only one positive root. 

An extensive numerical exploration for reasonable values of the parameters
has never yielded more than three positive real roots for $\mathcal{P}$.
Neither we found numerical evidence of limit cycles or deterministic
chaos. We therefore are confident that the bifurcation diagrams shown
in Figure \ref{fig:Bifurcations} determine all the qualitative dynamics
of the model equations (\ref{eq:nondimensional_model}).

\subsection{\label{sub:Pade_equilibria}Approximate explicit expressions for
the equilibria near the state of healthy vineyard}

For $\epsilon\ll1$, explicit, approximate expressions for the equilibria
of the model (\ref{eq:nondimensional_model}) may be sought perturbatively,
assuming an expansion of the form 
\begin{eqnarray*}
S(\epsilon) & = & \mathcal{D}+\epsilon S_{1}+\epsilon^{2}S_{2}+\epsilon^{3}S_{3}+\cdots\\
L(\epsilon) & = & \epsilon L_{1}+\epsilon^{2}L_{2}+\epsilon^{3}L_{3}+\cdots\\
I(\epsilon) & = & \epsilon I_{1}+\epsilon^{2}I_{2}+\epsilon^{3}I_{3}+\cdots\\
G(\epsilon) & = & \epsilon G_{1}+\epsilon^{2}G_{2}+\epsilon^{3}G_{3}+\cdots
\end{eqnarray*}
which represents a small correction upon the healthy vineyard equilibrium.
The perturbative analysis reveals that $S_{1}=L_{1}=I_{1}=G_{1}=0$.
That is, weak hotbeds at first perturbative order have no effect on
a healthy vineyard. The second and higher orders are non-zero, and
the information that they carry is best conveyed by using Padé approximants.
The (2,1) Padé approximation of the equilibrium computed with the
perturbative expansion up to the third order is the following
\begin{equation}
\begin{array}{l}
S(\epsilon)=\mathcal{D}-{\displaystyle \frac{\left(c_{2}c_{3}+c_{4}\left(c_{1}+c_{2}+c_{3}\left(c_{2}+1\right)\right)\right)}{c_{2}c_{4}\left(c_{1}+c_{3}-2\epsilon\mathcal{D}\right)}}\epsilon^{2}\\
L(\epsilon)={\displaystyle \frac{\left(c_{1}+c_{3}\right)\mathcal{D}}{c_{2}\left(c_{1}+c_{3}-2\epsilon\mathcal{D}\right)}}\epsilon^{2}\\
I(\epsilon)\,={\displaystyle \frac{\mathcal{D}}{\left(c_{1}+c_{3}-2\epsilon\mathcal{D}\right)}}\epsilon^{2}\\
G(\epsilon)={\displaystyle \frac{c_{3}\mathcal{D}}{c_{4}\left(c_{1}+c_{3}-2\epsilon\mathcal{D}\right)}}\epsilon^{2}
\end{array}\label{eq:Pad=0000E9}
\end{equation}

\subsection{\label{sub:Approx_saddle_node}Approximate position of the saddle-node
bifurcation}

If the vineyard's desired density $\mathcal{D}$ is sufficiently high,
for $\epsilon=0$ there are three equilibria: the healthy vineyard
stable node (\ref{eq:healthy vineyard}) (with no infected plants),
a saddle (with an intermediate number of infected plants), and another
stable node (with a high number of infected plants). As the parameter
$\epsilon$ grows, the branch of stable nodes which passes through
(\ref{eq:healthy vineyard}) and the branch of saddles move close
to each other, and meet in a saddle-node bifurcation at $\epsilon_{fold}$
(e.g. Figure \ref{fig:Bifurcations}(D) for $\mathcal{D}=6$). The
value of $\epsilon_{fold}$ and of the corresponding equilibrium value
of infected plants $I_{fold}$ may be approximated with explicit expressions,
as shown in Figure \ref{fig:fold_approximations}.

First we observe that for positive $\epsilon$ and sufficiently large
$\mathcal{D}$ the polynomial $\mathcal{P}$ has one real positive
root of size $O(\mathcal{D})$. The other roots, as $\mathcal{D}\to\infty$,
tend to the solutions of 
\[
I^{4}+c_{4}I^{2}+c_{4}\epsilon^{2}+\epsilon^{4}=0
\]
(where we have used the expressions (\ref{eq:coefficients}) divided
by $\mathcal{D}$). But this polynomial does not have real solutions.
Therefore we conclude that for fixed $\epsilon>0$ and asymptotically
large $\mathcal{D}$, the polynomial $\mathcal{P}$ has only one real
root, which is positive.

For $\epsilon=0$ the polynomial $\mathcal{P}$ has the root $I=0$.
The other equilibria are given by the solutions of 
\begin{equation}
\frac{q_{5}}{\mathcal{D}}I^{4}-c_{2}I^{3}+\frac{q_{3}}{\mathcal{D}}I^{2}-c_{2}c_{4}I+\frac{c_{2}c_{4}\left(c_{1}+c_{3}\right)}{\mathcal{D}}=0\label{eq:P4}
\end{equation}
where $q_{5},q_{3},q_{1}$ are given by (\ref{eq:coefficients}) with
$\epsilon=0$. For $\mathcal{D}\to\infty$ one of the solutions of
(\ref{eq:P4}) approaches zero. Therefore it may be approximated by
neglecting the terms of order higher than the first, yielding
\begin{equation}
\left.I\right|_{\epsilon=0}\approx\frac{c_{1}+c_{3}}{\mathcal{D}}.\label{eq:Iepsi0}
\end{equation}

For $0\le\epsilon\le\epsilon_{fold}$, a smooth family of equilibria
connects the equilibrium corresponding to (\ref{eq:Iepsi0}) to the
healthy vineyard equilibrium (\ref{eq:healthy vineyard}), changing
stability at $\epsilon_{fold}$. But for $\mathcal{D}\to\infty$ it
must be $\epsilon_{fold}\to0$ because for large $\mathcal{D}$ and
positive $\epsilon$, $\mathcal{P}$ has only one real solution. Thus,
if the family of equilibria is a smooth curve, asymptotically for
large $\mathcal{D}$, it must be $0<I_{fold}<\left.I\right|_{\epsilon=0}$
and $\epsilon_{fold}\propto I_{fold}$. We have verified numerically
for a large number of values of $c_{1},$ $c_{3}$ and $\mathcal{D}$,
that 
\[
I_{fold}=\epsilon_{fold}=\frac{\left.I\right|_{\epsilon=0}}{4}
\]
is a very good approximation for the position of the saddle-node bifurcation,
except for the values of $\mathcal{D}$ so low that for $\epsilon=0$
, $\mathcal{P}$ has only the real root $I=0$ (e.g. the case $\mathcal{D}=4.5$
in Figure \ref{fig:Bifurcations}(D)).

\end{document}